\documentstyle[aps,prbbib,twocolumn,epsf]{revtex}
\begin{document}
\medskip
\draft
\title{Heat current in parametric quantum pump}

\author{Baigeng Wang $^1$ and Jian Wang$^{1,2}$}

\address{1. Department of Physics, The University of Hong Kong, 
Pokfulam Road, Hong Kong, China\\
2. Institute of Solid State Physics, Chinese Academy of Sciences,
Hefei, Anhui, China}
\maketitle

\begin{abstract}
We investigate the heat flow in the parametric quantum pump. Using
the time dependent scattering matrix theory, we have developed a 
general theory for the pumped heat current at finite pumping 
amplitude and frequency. We have applied our theory to a double
barrier structure and studied pumped heat current in both the weak
and strong pumping regimes as different system parameters vary. By 
comparing the pumped heat current and the power of Joule heat generated 
in the system, we found that the double barrier structure can
function as an optimal pump in the strong pumping regime.

\end{abstract}


The physics of adiabatic quantum pump has attracted great attention
recently\cite{brouwer,switkes,zhou,wagner,Avron,aleiner1,wei1,vavilov,brouwer2,sharma,kravtsov,aleiner,shutenko,levinson,buttiker1,wbg1,wang1,wang2}.
The quantum pump is realized by varying the geometric parameters 
of the quantum dot, by which the current is generated. At the same time, 
heat current is also produced and accompanied with the dissipation. Avron
et al\cite{avron} have given a lower bound for the dissipation in a
quantum channel which is defined as the difference between the 
heat current and the power of Joule heat. The pump is optimal if the
heat current equals to the power of Joule heat\cite{avron}. As a result, 
the optimal pump is noiseless and charge transported is quantized. 
In a recent paper, Moskalets and Buttiker\cite{buttiker} have also 
considered the dissipation in an adiabatic quantum pump. The heat current 
and the noise
has been formulated in terms of a parametric emissivity matrix. The theory 
of Moskalets and Buttiker is in the weak pumping regime (quadratic order 
in pumping amplitude) and can go to the finite frequency if one goes 
beyond the hypothesis of instant scattering. In this paper, we develop 
a general theory for the heat current which is valid for finite pumping 
amplitude and finite frequency. This allows us to study the heat
current in both weak and strong pumping regimes. Our theory is based 
on the time dependent scattering matrix theory\cite{buttiker,vavilov} 
and goes beyond the instant scattering hypothesis. We have applied our 
theory to a double barrier structure and studied pumped heat current 
as different system parameters such as Fermi energy, pumping amplitude, 
and phase difference vary. In the weak pumping regime, the pumped heat 
current increases quadratically as pumping amplitude increases. The
dependence becomes linear in the strong pumping regime. In the
strong pumping regime, the pumped heat current shows strong
nonlineality as a function of phase difference between two pumping
potentials. The heat current for single pump has also been studied.
We found that the amplitude of the pumped heat current for the
single pump is of the same order as that of two pumping potentials.
Recently there is a concern of whether a genuine optimal pump with
nonvanishing transmission coefficient exists\cite{alekseev}. In this
paper, we give a nontrivial example of optimal pump. By comparing 
the pumped heat current and the power of Joule heat 
generated in the system, we found that the double barrier structure 
is an optimal pump in the strong pumping regime.

We start with the general definition for the heat current in scattering 
matrix theory\cite{buttiker}, 

\begin{equation}
I_{q,\alpha} = \lim_{\Delta t -> \infty}
\frac{1}{\Delta t}\int^{\Delta t}_0 dt <{\hat
I}_{q,\alpha}>
\end{equation}
where the heat current operator is ${\hat I}_{q,\alpha} = {\hat
I}_{E,\alpha} - E_F {\hat I}_{e,\alpha}$ and $<...>$ denotes the 
quantum average.  Here ${\hat I}_{E,\alpha}$
is the energy current operator given by
\begin{equation}
{\hat I}_{E,\alpha} = -i[ \partial_t {\hat
b}^{\dagger}_\alpha(t) {\hat b}_\alpha(t) - \partial_t {\hat
a}^\dagger_\alpha(t) {\hat a}_\alpha(t)]
\label{heat1}
\end{equation}
and ${\hat I}_{e,\alpha}$ is the electric current operator,
\begin{equation}
{\hat I}_{e,\alpha} = {\hat
b}^{\dagger}_\alpha(t) {\hat b}_\alpha(t) - {\hat
a}^\dagger_\alpha(t) {\hat a}_\alpha(t)
\end{equation}
where the operators ${\hat b}_\alpha$ and ${\hat a}_\alpha$ are 
annihilation operators for the outgoing and incoming carriers in the 
lead $\alpha$. They are related by the scattering matrix,

\begin{equation}
{\hat b}_\alpha(t) = \sum_\beta \int dt' s_{\alpha \beta} (t,t')
{\hat a}_\beta(t')
\label{bb}
\end{equation}
where the time-dependence of the scattering matrix is
due to the slowly time-varying pumping potential $X(t)$. 
The distribution function can be obtained by taking the quantum 
average,\cite{buttiker} 
\begin{equation}
<{\hat a}^\dagger_\alpha(E) {\hat a}_\beta(E')> = \delta_{\alpha \beta}
\delta(E-E') f_\alpha(E)
\label{fermi}
\end{equation}
where ${\hat a}_\alpha(E)$ is the Fourier transform of ${\hat a}_\alpha(t)$ 
and $f(E)$ is the Fermi distribution function. For the purpose of 
presentation, we calculate the energy current first and the electric 
current can be calculated in a similar fashion.  From Eqs.(\ref{heat1}), 
(\ref{bb}), and (\ref{fermi}), the energy current is given by

\begin{eqnarray}
I_{E,\alpha} &=& -\lim_{\Delta t -> \infty}
\frac{i}{\Delta t}\int^{\Delta t}_0 dt \int dt_1
dt_2 \sum_\beta s_{\alpha \beta}(t,t_1) \nonumber \\
&& f(t_1-t_2) \partial_t s^\dagger_{\alpha \beta}(t,t_2) 
- \int \frac{dE}{2\pi} E f(E)
\label{heat2}
\end{eqnarray}
where $f(t) \equiv \int (dE/2\pi) \exp(-iEt) f(E)$. Now we will focus 
on the first term (denoted as $I^{(1)}_{E,\alpha}$) in Eq.(\ref{heat2}).
After changing of the variable $t_0=(t_1+t_2)/2$ and $\tau=t_1-t_2$
and using the following Wigner transform for the scattering
matrix\cite{vavilov},

\begin{equation}
s(t,t') = \int \frac{d\epsilon}{2\pi} e^{-i\epsilon (t-t')} 
s(\epsilon,\frac{t+t'}{2})
\end{equation}
Eq.(\ref{heat2}) becomes,

\begin{eqnarray}
&&I^{(1)}_{E,\alpha} = 
\lim_{\Delta t -> \infty}
\frac{-i}{4\pi^2\Delta t}\int^{\Delta t}_0 dt \int dt_0
d\tau d\epsilon_1 d\epsilon_2 
~ f(\tau) 
\nonumber \\
&& e^{-i\epsilon_1(t-t_0-\tau/2)}
e^{i\epsilon_2(t-t_0+\tau/2)} 
\sum_\beta s_{\alpha \beta}(\epsilon_1,
\frac{t+t_0}{2} +\frac{\tau}{4}) \nonumber \\
&& [\partial_t s^\dagger_{\alpha \beta}(\epsilon_2,\frac{t+t_0}{2}
-\frac{\tau}{4}) + i\epsilon_2 s^\dagger_{\alpha
\beta}(\epsilon_2,\frac{t+t_0}{2}-\frac{\tau}{4})]
\nonumber \\
\end{eqnarray}
Changing the variables again to $\tau_1=t-t_0$ and $t'=(t+t_0)/2$ and
integrating over $\tau_1$, we obtain

\begin{eqnarray}
&&I^{(1)}_{E,\alpha} = 
\lim_{\Delta t -> \infty}
\frac{-i}{2\pi\Delta t}\int^{\Delta t}_{-\Delta t} dt' 
d\tau d\epsilon \sum_\beta s_{\alpha \beta}(\epsilon,t'+\tau/4) f(\tau) 
\nonumber \\
&&[(1/2) \partial_{t'} s^\dagger_{\alpha \beta}(\epsilon,t'-\tau/4) 
+ i\epsilon s^\dagger_{\alpha \beta}(\epsilon,t'-\tau/4)]
~ e^{i\epsilon \tau}
\label{i1}
\end{eqnarray}
To get the heat current up to $\omega^2$, it is enough to expand 
$s_{\alpha \beta}$ up to the second order in $\tau$. We obtain

\begin{eqnarray}
I_{E,\alpha} &=& -\frac{1}{8\pi T_p}\int^{T_p}_0 dt 
\int dE [2\partial_E f -E \partial^2_E f] \nonumber \\
&& \sum_\beta \partial_t s_{\alpha \beta}(E,t) 
\partial_t s^\dagger_{\alpha \beta}(E,t) \nonumber \\
&-& \frac{1}{2\pi T_p}\int^{T_p}_0 dt 
\int dE ~ E ~\partial_E f \nonumber \\
&& \sum_\beta {\rm Im} [\partial_t s^\dagger_{\alpha \beta}(E,t) 
s_{\alpha \beta}(E,t)] 
\end{eqnarray}
where the zeroth order term in $\tau$ in Eq.(\ref{i1}) has been canceled 
by the second term in Eq.(\ref{heat2}) and $T_p$ is the period of
the pumping cycle. Including the electric current, the heat current is 
given by
\begin{eqnarray}
I_{q,\alpha} &=& -\frac{1}{8\pi T_p}\int^{T_p}_0 dt 
\int dE [2\partial_E f -(E-E_F) \partial^2_E f] \nonumber \\
&& \sum_\beta \partial_t s_{\alpha \beta}(E,t) 
\partial_t s^\dagger_{\alpha \beta}(E,t) \nonumber \\
&-& \frac{1}{2\pi T_p}\int^{T_p}_0 dt 
\int dE (E-E_F) \partial_E f \nonumber \\
&& \sum_\beta {\rm Im} [\partial_t s^\dagger_{\alpha \beta}(E,t) 
s_{\alpha \beta}(E,t)] 
\label{final}
\end{eqnarray}
Note that the second term in Eq.(\ref{final}) vanishes at zero temperature. 
In the adiabatic regime, we have\cite{vavilov} $\partial_t s_{\alpha 
\beta} = \sum_i[\partial_{X_i} s_{\alpha \beta} \partial_t X_i + 
\partial_{{\dot X}_i} s_{\alpha \beta} \partial_t {\dot X}_i + ...]$
where ${\dot X} \equiv dX/dt$. 
Up to the order $\omega^2$, we can neglect the contribution from 
$\partial_{{\dot X}_i} s_{\alpha \beta}$. 
At zero temperature, Eq.(\ref{final}) becomes,

\begin{eqnarray}
I_{q,\alpha} &=& \frac{3}{8\pi T_p}\int^{T_p}_0 dt 
\sum_\beta \sum_{ij} \partial_{X_i} s_{\alpha \beta}  \nonumber \\
&&~ \partial_{X_j} s^\dagger_{\alpha \beta} ~ \partial_t X_i
~ \partial_t X_j
\label{final1}
\end{eqnarray}
It is straightforward to obtain the heat current at higher order 
in frequency. To do that, we have to keep the higher order
term in the expansion of $\tau$ in Eq.(\ref{i1}) and include the
contribution of higher order derivatives\cite{zhou,vavilov} such as
${\dot X}$. For instance, expanding Eq.(\ref{i1})
to the third order in $\tau$ and including the contribution from the
electric current, we obtain (up to $\omega^3$),

\begin{eqnarray}
I^{(3)}_{q,\alpha} &=& \frac{1}{12\pi T_p}\int^{T_p}_0 dt 
\int dE \partial^2_E f \sum_\beta {\rm Im} \nonumber \\
&&[\partial^2_t s_{\alpha \beta} \partial_t s^\dagger_{\alpha \beta}]
\label{final2}
\end{eqnarray}
There are also third order corrections from Eq.(\ref{final}) where 
the term $\partial_{{\dot X}_i} s_{\alpha \beta}$ should be kept.
We now consider the limiting case of Eq.(\ref{final1}) when the 
pumping amplitude is small. 
For two probe pumping: $X_1(t) = X_1 \sin(\omega t)$ and $X_2(t) = 
X_2 \sin(\omega t +\phi)$, we see that the lowest order pumping 
amplitude in Eq.(\ref{final1}) is quadratic by neglecting $X$
dependence in $\partial_X s_{\alpha \beta}$. In this case, it is 
straightforward to show that Eq.(\ref{final1}) is reduced to 

\begin{eqnarray}
I_{q,\alpha} &=& \frac{3\omega}{16\pi} [
X_1^2 \sum_\beta |\partial_{X_1} s_{\alpha \beta}|^2  
+X_2^2\sum_\beta |\partial_{X_2} s_{\alpha \beta}|^2  
\nonumber \\
&+& 2\cos\phi ~ X_1 X_2 \sum_\beta {\rm Re} (\partial_{X_1} s_{\alpha \beta}
\partial_{X_2} s^\dagger_{\alpha \beta})]
\label{weak}
\end{eqnarray}
which agrees\cite{foot1} with the result of Ref.\onlinecite{buttiker}.

We now apply our formula Eq.(\ref{final1}) to a one-dimensional
quantum structure which modeled by double barrier potential 
$U(x)=X_1 \delta (x+a)+X_2 \delta (x-a)$ where $2a$ is the well width. 
For this system the Green's function $G(x,x')$ can be calculated 
exactly\cite{yip}. With $G(x,x')$ we can calculate scattering matrix 
from the Fisher-Lee relation\cite{lee} $s_{\alpha\beta}=
-\delta_{\alpha \beta}\,+\,i\hbar v G(x_\alpha,x_\beta)$, with $v$ 
the electron velocity in the lead.  The adiabatic pump that we consider 
is operated by changing barrier heights adiabatically and periodically: 
$X_1=V_{10}+V_p\sin(\omega t)$ and $X_2=V_{20}+V_p\sin(\omega t+\phi)$.  
This can be achieved by microfabricating metallic gates at the barrier 
region and applying a time dependent gate potential. Since the pumped 
current is proportional to $\omega^2$, we will set $\omega=1$ for 
convenience. Finally the unit is set by $\hbar=2m=1$\cite{foot2}.

We first study the pumped heat current with two pumping potentials.
Fig.1 depicts the pumped heat current exiting from left lead versus 
Fermi energy for the symmetric barriers at small pumping 
amplitude. We have also plotted the transmission 
coefficient (solid line) versus Fermi energy for comparison. 
The physical picture of heat flow suggested by Moskalets and 
Buttiker\cite{buttiker} is the following: as electron scattered by the 
oscillating barriers or scatterers, the absorption of energy quantum 
$\hbar\omega$ creates an electron-hole pair.  The flow of electron-hole 
pair leads to the heat transfer. In the symmetric case, we have 
$I_{q,L}=I_{q,R}$. We see that the heat current is peaked at the resonant 
levels and is clearly proportional to the density of states of the 
scattering region. At phase difference $\phi=3\pi/4$ (short-dashed line),
the lineshape of $I_q$ is approximately Lorentzian similar to that
of transmission coefficient. As $\phi$ decreases, the lineshape
is broadened and deviates from Lorentzian considerably. For large
pumping amplitude (see inset of Fig.1), we see that there is significant
heat current even in the off resonant case (when Fermi enegy is not in
line with the resonant level in the static case when pumping
potential is off). This is because in the strong pumping regime, the 
instantaneous resonant level oscillates with a large amplitude and hence 
can generate heat current in a broad range of energy. Up to the order
of $\omega^2$, the heat currents $I_{q,L}$ and $I_{q,R}$ are all
positive and flow from the scatterers to the reservoir. It is no
longer true when high order frequency contribution is included.
Fig.2 shows the heat current as a function of phase difference for
different pumping amplitudes. In general, there are two extreme points 
for the heat current at $\phi=0$ and $\phi=\pi$, where the former 
corresponds to maximum in heat current and latter for the minimum.  
In the weak pumping regime, the sinusoidal behavior is seen. In the 
strong pumping regime, however,  
we see significant derivation from the sinusoidal behavior. As a result, 
when increasing $\phi$ from $\phi=0$ to $\pi$, the heat current drops 
not as fast as in the weak pumping regime away from $\phi=\pi$. 
However, near $\phi=\pi$, the heat current decreases much faster. 
Fig.3 displays the heat current as a function of pumping amplitude.
The general behaviors of heat current at different $\phi$ are similar.
We see that initially, the heat current increases quadratically with 
pumping amplitude in the weak pumping regime and then quickly approaches 
to the linear regime in the strong pumping regime. This behavior persists 
for single pumping potential. In the inset, we present the result 
for asymmetric barriers.  Here $v_{10}=4$, $v_{20}=6$, and $v_p=0.5$, 
$I_{q,L}$ is about twice as large as $I_{q,R}$. This is reasonable 
because it is easier for heat current to tunnel through the lower 
barrier. In general, the electric pumped current is linear in frequency 
for two pumping potentials whereas for single pumping potential, it must 
be zero up to the first order in frequency\cite{vavilov,wbg1}.
The heat current is different. For both single pump and two potential 
pump, the heat currents are of order $\omega^2$. In the weak pumping 
limit, there is a simple relationship between the following three 
heat currents\cite{buttiker}: $I_{\pi/2}=I_L+I_R$. where $I_{\pi/2}$ 
is the heat current of two pumping potentials with phase difference 
$\phi=\pi/2$, $I_L$ is the heat current of single pump due to the left 
oscillating barrier, and $I_R$ heat current due to the right barrier. 
In the strong pumping regime, the scattering matrix depends on both $X_1$ 
and $X_2$ in a nonlinear fashion, this simple relation is no longer valid. 
Denoting the ratio $\tau \equiv I_{\pi/2}/(I_L+I_R)$, we found that
at small pumping amplitude, $\tau \sim 1$. As one increases the pumping 
amplitude the ratio $\tau$ decreases and quickly saturates around 
$\tau \sim 0.8$. This is understandable since the behavior of $I_{\pi/2}$, 
$I_L$, and $I_R$ are similar as one increases the pumping amplitude, 
one obtains the nearly constant ratio in the linear pumping amplitude 
dependence regime (see Fig.3). 

Finally, we study the dissipation of the pump by comparing the heat 
current and the power of Joule heat produced by the electric 
current\cite{foot3}. We found that the heat current is always larger 
than the corresponding power of Joule heat as predicted by 
Ref.\onlinecite{avron}. In Fig.4, we plot the pumped charge per cycle and
the ratio between the power of Joule heat and the heat current as a 
function 
of pumping amplitude $v_p$. We see that the pumped charge increases as 
$v_p$ increases. In the large $v_p$ limit, the pumped charge will 
eventually reach the maximum value $Q=e$\cite{levinson,wang1}. We see 
from Fig.4 that the ratio also approaches to one just like the pumped 
charge. Therefore, we conclude that the double barrier pump we studied 
can be optimal in very strong pumping limit.

In summary, we have developed a general theory for the pumped heat current 
at finite pumping amplitude and frequency which allows us to investigate 
the pumped heat current in both weak and strong pumping regimes as 
different system parameters vary. As the pumping amplitude varies, we 
observed
the crossover of pumped heat current from quadratic dependence in the 
weak pumping regime to linear dependence in the strong pumping regime. 
In the strong pumping regime, the pumped heat current shows strong
nonlineality as a function of phase difference between two pumping
potentials. For the single pump, we found that amplitude of pumped
heat current is of the same order as that of two pumping potentials.
Finally, our numerical results show that the double barrier structure 
we examined can be an optimal pump in the strong pumping regime.

\section*{Acknowledgments}
We gratefully acknowledge support by a RGC grant from the 
SAR Government of Hong Kong under grant number HKU 7091/01P.

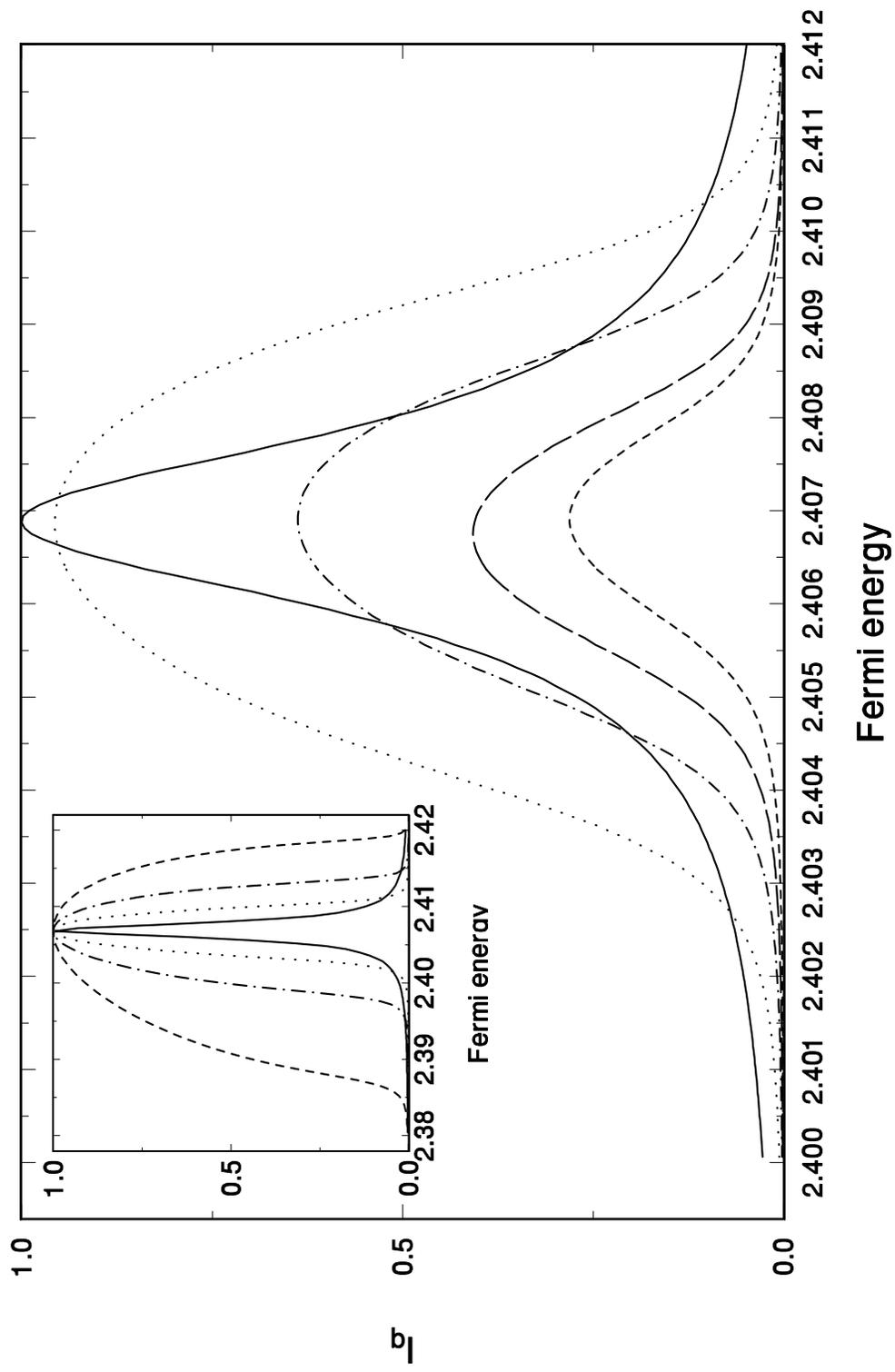
\begin{figure}
\caption{
The transmission coefficient (solid line) and heat current as a
function of Fermi energy at different phase differences between two
pumping potentials: $\phi=0$ (dotted line), $\phi=\pi/2$ (dot-dashed 
line), $\phi=3\pi/4$ (short-dashed line).  The heat current for a single 
pump (long-dashed line).  For illustrating purpose, the heat current 
has been multiplied by 10/3. Other parameters: $v_{10}=v_{20}=80$ and 
$v_p=4$. Inset: the heat current as a function of Fermi energy at 
different pumping amplitude: $v_p=5$ (dotted line), $v_p=10$ (dot-dashed 
line), $v_p=20$ (dashed line). The transmission coefficient is also shown 
(solid line). The heat current has been normalized to one. The scaling
factors are: 8/3 for $v_p=5$, 4/3 for $v_p=10$, and 2/3 for $v_p=20$.
Other parameters: $v_{10}=v_{20}=80$, $\phi=0$. 
}
\end{figure}

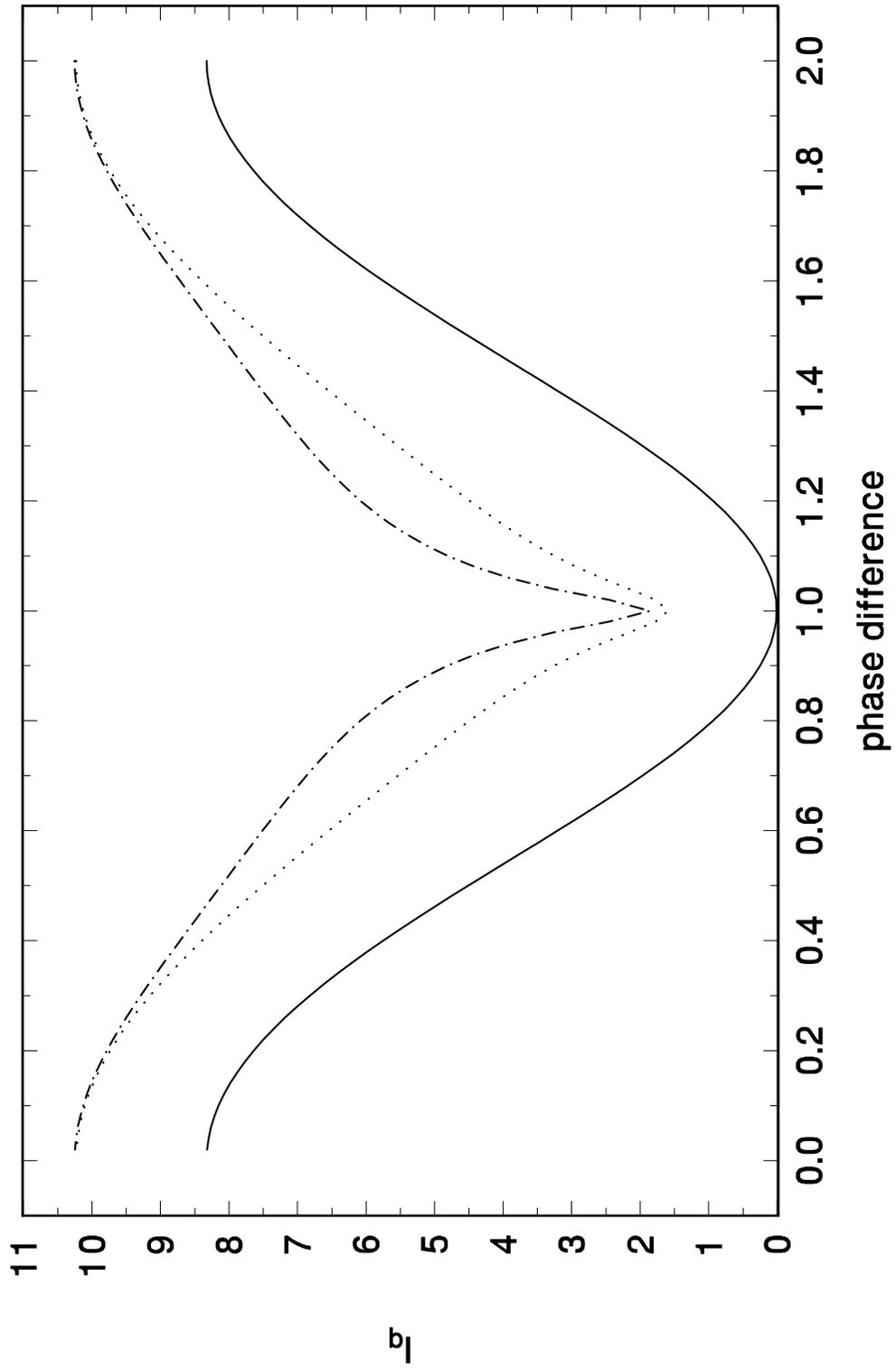
\begin{figure}
\caption{
The heat current as a function of phase differences $\phi/\pi$
at different
pumping amplitudes: $v_p=1$ (solid line), $v_p=20$ (dotted line), 
$v_p=40$ (dotted-dash line). The heat current for a single pump 
(long-dashed line). For illustrating purpose, the heat current has 
been multiplied by a factor of 200, 20/3, 10/3, respectively for $v_p=1, 
20, 40$. Other parameters: $v_{10}=v_{20}=80$ and $E_F=2.4069$.
}
\end{figure}

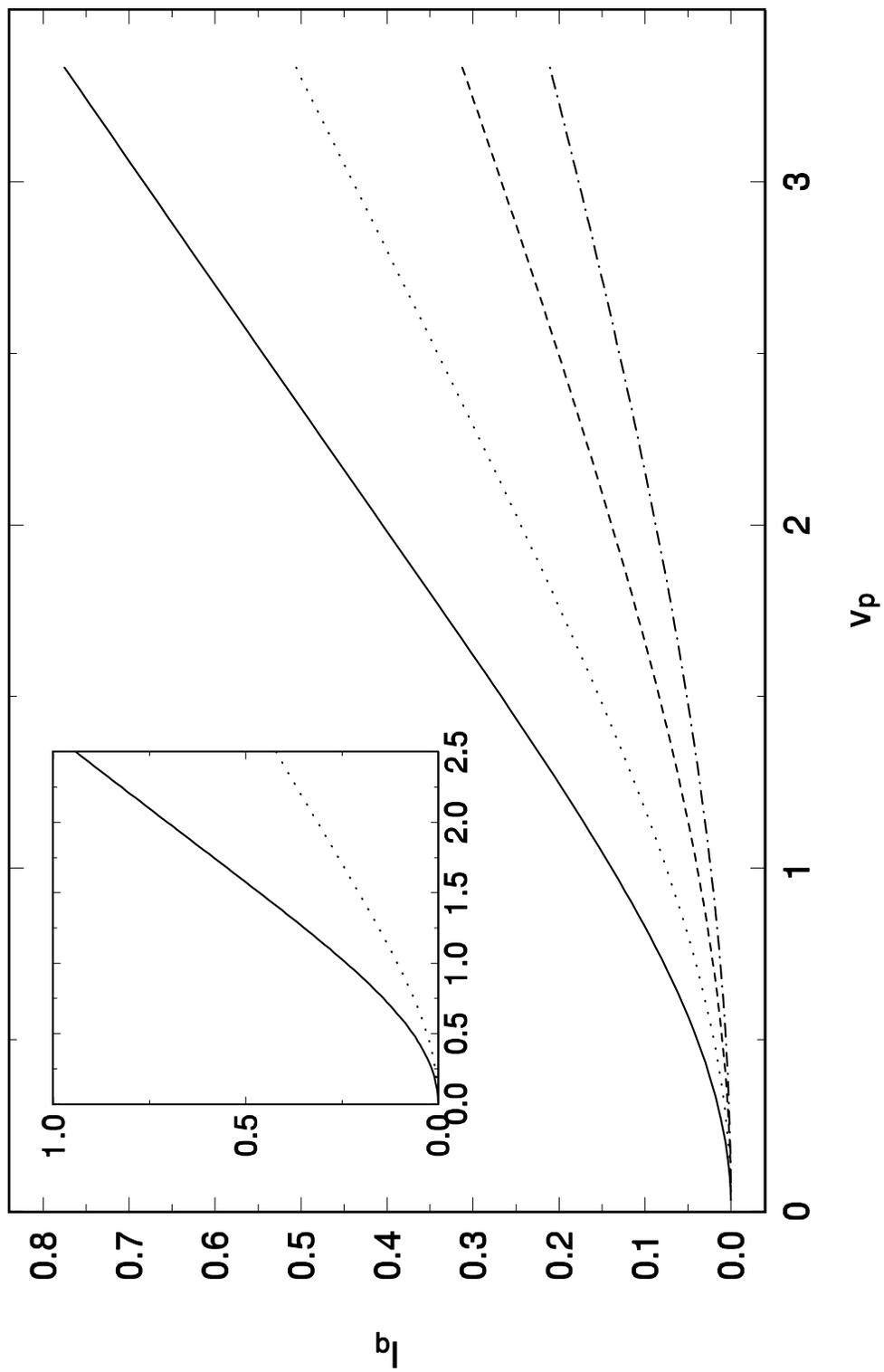
\begin{figure}
\caption{
The heat current as a function of pumping amplitude at different
phase difference: $\phi=0$ (solid line), $\phi=\pi/2$ (dotted line), 
$\phi=3\pi/4$ (dotted-dash line). The heat current for a single pump 
(dashed line).  The heat current has been multiplied by a
factor of 10/3. Other parameters are the same as the inset of Fig.2.
Inset: $I_{q,L}$ (solid line) and $I_{q,R}$ (dotted line) versus 
pumping amplitude for asymmetric barriers. Here $\phi=0$ and $E_F=1.65$. 
The scaling factor is 1/3.
}
\end{figure}

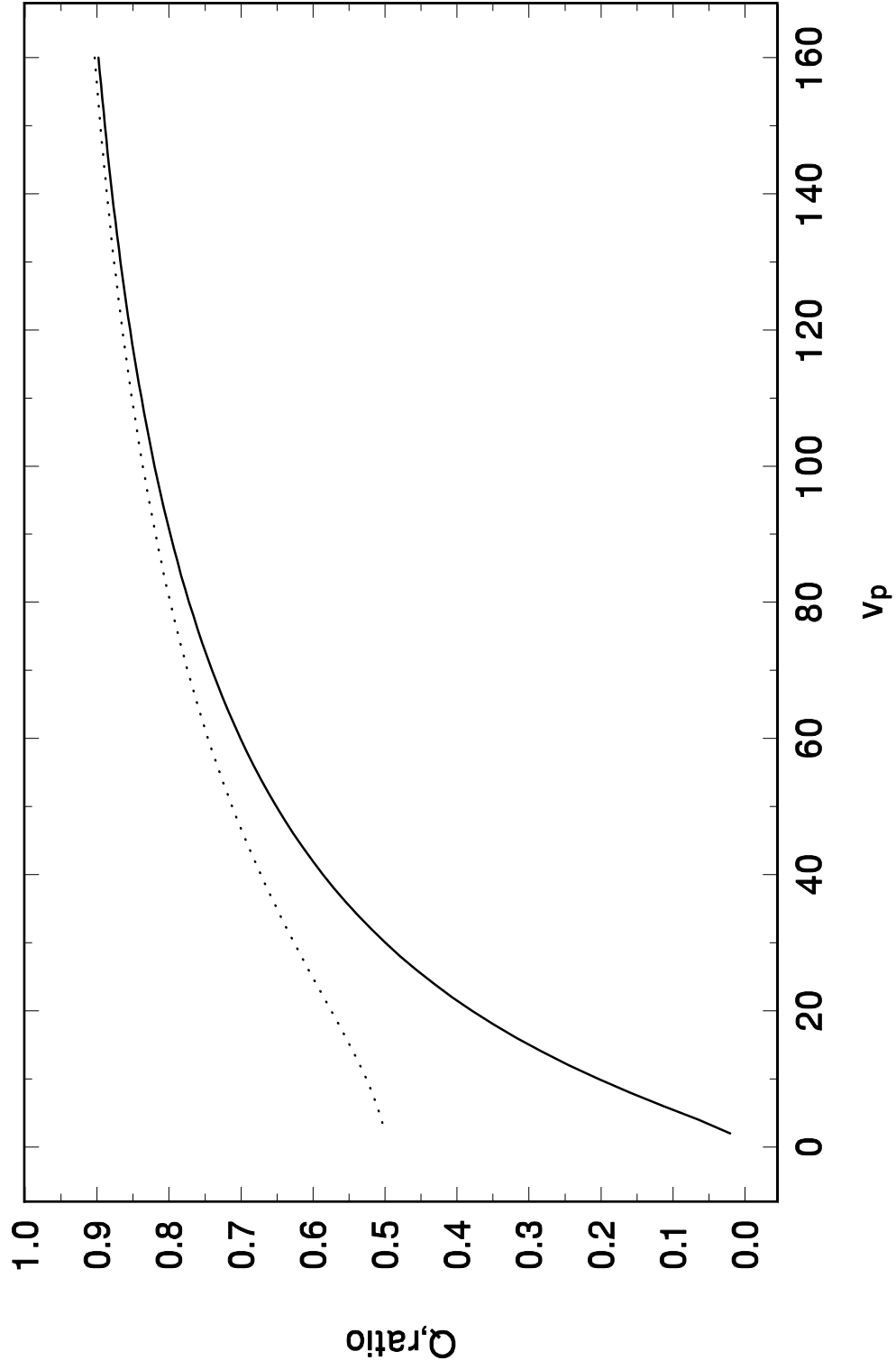
\begin{figure}
\caption{
The pumped charge per cycle (solid line), the ratio between the power 
of Joule heat and the heat current (dotted line) as a function of pumping 
amplitude. The system parameters are $v_{10} = v_{20}=80$, $\phi=3/pi/4$,
and $E_F=2.4069$. 
}
\end{figure}


\begin{thebibliography}{00}
\bibitem{foot1}
The difference between our Eq.(\ref{final}) and Eq.(21) in 
Ref.\onlinecite{buttiker} is due to the fact that the scattering matrix 
and hence $T_{\pm \omega}$ defined in Ref.\onlinecite{buttiker} is energy 
independent. As a result, a term $\omega^2 (E-E_F) ~\partial_E f ~
(\partial_E T_{-\omega} +\partial_E T_{+\omega})$ is missing in Eq.(21) 
of Ref.\onlinecite{buttiker}. This can be easily understood from Eq.(6) 
of Ref.\onlinecite{wbg1}. Finally, in order to compare our result 
with that of Ref.\onlinecite{buttiker} one should multiply $1/4$ in 
Eqs.(21) and (39) of Ref.\onlinecite{buttiker}.
\bibitem{foot2}
For the GaAs system with $a=1000A$, the energy uint is $E=56 \mu eV$.
\bibitem{foot3}
The power of Joule heat is defined as $(3\pi/2T_p) \int_0^{T_p} dt 
(dQ/dt)^2$ where $Q = \int_0^{T_p} dt (dQ/dt)$ gives the pumped charge
and $I_p = Q/T_p$ the pumped electric current.
\end{thebibliography}
\end{document}